\documentclass[apj]{emulateapj} 
\usepackage{graphicx}
\usepackage{epstopdf}
\usepackage{color}

\newcommand{\beq}{\begin{equation}}
\newcommand{\eeq}{\end{equation}}

\newcommand{\Msunh}{\>h^{-1}{\rm M}_\odot}
\newcommand{\Msun}{\>{\rm M}_\odot}

\shorttitle{Classical bulges, supermassive blackholes and AGN feedback}
\shortauthors{Lu and Mo}

\begin{document}


\title{Classical bulges, supermassive blackholes and AGN feedback:
         Extension to low-mass galaxies}

\author{Zhankui Lu\altaffilmark{1} and H.J. Mo\altaffilmark{1,2}}

\altaffiltext{1}{Department of Astronomy, University of Massachusetts, Amherst
MA 01003-9305}
\altaffiltext{2}{Department of Astronomy, University of Science and
  Technology, Hefei, Anhui 230026, China}


\begin{abstract}
The empirical model of \citet{LuZ14a} for the relation 
between star formation rate and halo mass growth is adopted 
to predict the classical bulge mass ($M_{\rm cb}$)
- total stellar mass ($M_\star$) relation for central galaxies. 
The assumption that the supermassive black hole 
(SMBH) mass ($M_{\rm BH}$) is directly proportional 
to the classical bulge mass, with the proportionality 
given by that for massive galaxies, predicts 
a $M_{\rm BH}$ - $M_\star$ relation that matches well 
the observed relation for different types of galaxies. 
In particular,  the model reproduces the strong transition at 
$M_\star=10^{10.5}$ - $10^{11}\Msun$, below which $M_{\rm BH}$ 
drops rapidly with decreasing $M_\star$. Our model
predicts a new sequence at $M_\star <10^{10.5}\Msun$, 
where $M_{\rm BH} \propto M_\star$ but the amplitude is 
a factor of $\sim 50$ lower than the amplitude of 
the sequence at $M_\star>10^{11}\Msun$. 
If all SMBH grow through similar quasar modes with
a feedback efficiency of a few percent, the energy produced 
in low-mass galaxies at redshift $z\ga 2$ can heat 
the circum-galactic medium up to a specific entropy 
level that is required to prevent excessive star formation 
in low-mass dark matter halos.      
\end{abstract}


\keywords{galaxies: formation -- galaxies: evolution}


\section{INTRODUCTION}

 The co-existence of supermassive black holes (hereafter SMBH) 
and early type galaxies, such as elliptical galaxies and the bulges 
of spiral galaxies, have been found ubiquitous \citep[e.g.][]{Kormendy95}.
Tight correlation has been observed between the SMBH mass 
and the stellar mass and velocity dispersion of the 
host spheroid  component. The SMBH mass is roughly proportional to
the spheroid mass, with a ratio between the two about 0.1\% 
\citep{McLure02,Haring04} to 0.5\% \citep[see][for a review]{Kormendy13}.

The observed scaling relation suggests that 
the growths of both the SMBH and the spheroid 
components are driven by similar physical processes.
The most popular scenario is mergers of galaxies, in 
which a large amount of cold gas originally supported by
a rotation disk can lose angular momentum and sink 
toward the center to fuel both the SMBH growth and the 
star formation. In addition, the energy feedback from the 
AGN associated with the SMBH accretion may heat and/or 
eject the cold gas, suppressing both star formation and 
SMBH growth \citep[e.g.][]{Springel05}.
Indeed, with plausible assumptions about
AGN feedback and about how star formation and SMBH growth 
are related to the cold gas density, numerical simulations and 
semi-analytical models can reproduce the general trend
in the observed scaling relation 
\citep[e.g.][]{Croton06,diMatteo08}.
In addition to the merger scenario, other mechanisms have also
been proposed, such as the direct accretion of low angular
momentum cold flow \citep{diMatteo12, Dubois12}, the intense 
gravitational instability in clumpy disks \citep{Bournaud11},
 and some secular processes \citep{Dubois14}.
However, it is not clear how much these mechanisms contribute
to the connection between SMBHs and their host galaxies.

Recent observations have extended the detections of 
SMBH and their mass measurements to late-type galaxies 
and to galaxies in the low-mass end
\citep[e.g.][]{Filippenko03,Greene04,Greene07,
Greene08,Jiang11a, Jiang11b}. 
These observations indicate a number of interesting 
deviations from the scaling relation derived from massive, 
early-type galaxies. First, it was found that many late-type 
spiral galaxies, which do not have a significant classical bulge 
component but contain a pseudo-bulge formed through 
secular evolution of the disk, do contain central SMBH. 
At a fixed bulge mass (luminosity) the SMBH masses are 
about a factor of 5 to 10 lower than expected from the 
correlation seen for classical bulges, and the scatter 
is also significantly larger \citep[see][]{Kormendy13}.
Or equivalently, the correlation between the central BH and 
the bulge is dramatically bent \citep{Graham12, Graham13, Scott13}.
Second, SMBH have also been observed in dwarf galaxies, such as 
dwarf spheroids and dwarf ellipticals 
\citep[e.g.][]{Barth04, Reines13}. It is still unclear
whether the SMBH in the late-type and low-mass 
galaxies have formed through the same processes 
as in classical bulges,  or through some completely 
different mechanisms. 

\begin{figure*}
\plotone{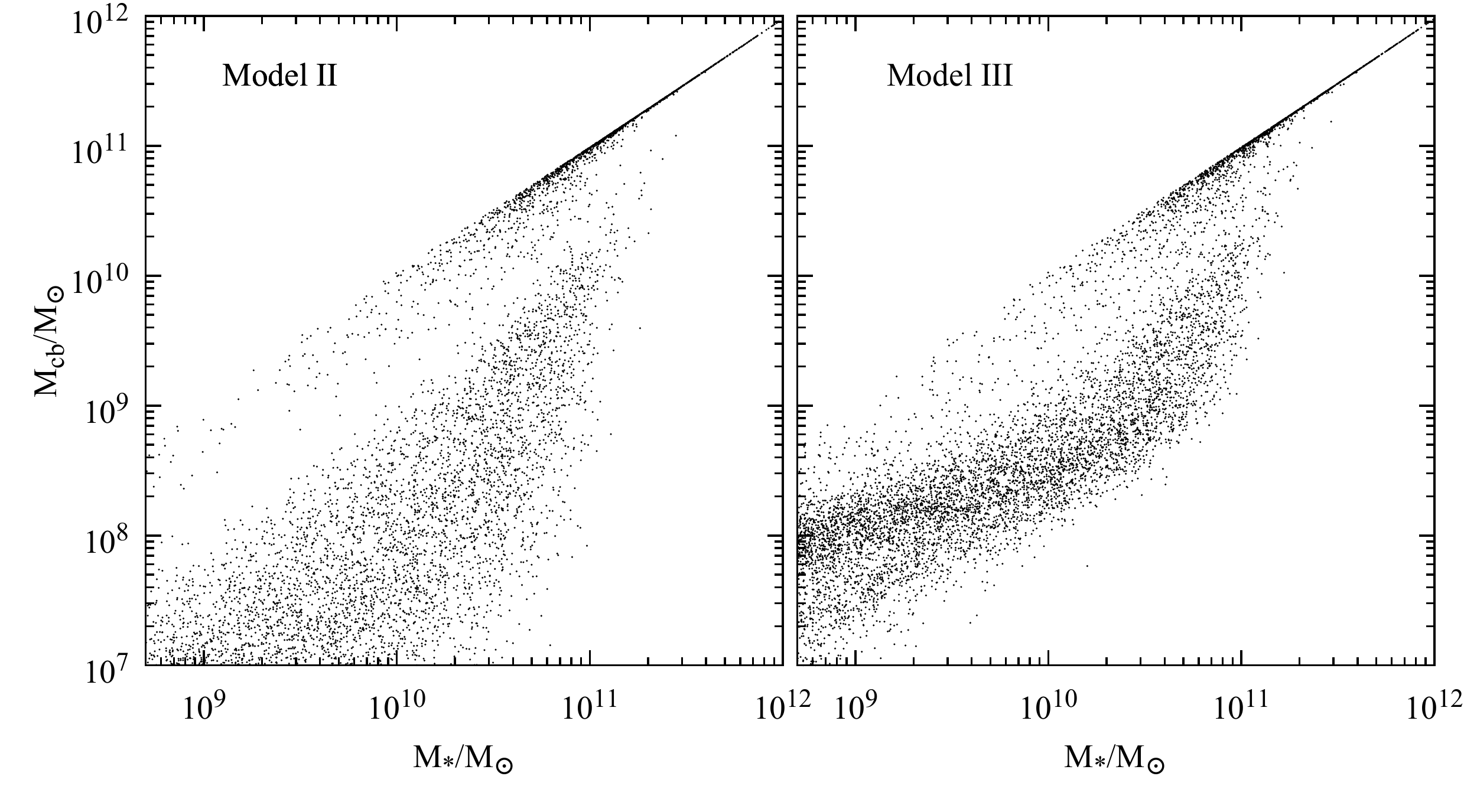}
\caption{
Relation between the mass of the classical bulge (formed via 
merger of galaxies)
and the total stellar mass of the host galaxy, predicted by Model II
(left) and Model III (right). 
}
\label{fig_bulge}
\end{figure*}

Recent modeling of \citet{LuZ14a,LuZ14b} of the observed 
populations of galaxies at different redshifts shows 
that all galaxies contain a classical bulge component
produced by major galaxy-galaxy merger.
The significance of such classical bulge depends 
strongly on the total stellar mass of the galaxy.  
In galaxies with sub-Galactic masses, this component is old 
in stellar population, small in stellar mass by an order 
of magnitude than pseudo-bulges \citep[e.g.][]{Fukugita98},  
and so may be missed in observation owing to 
the presence of other dominating 
components. In this paper, we demonstrate that 
the SMBH hosted by late-type galaxies can also be 
explained by the co-evolution with the classical bulge 
components potentially hidden in these galaxies. In \S\ref{sec_model}, 
we predict the relation between the classical bulge mass 
and total stellar mass of galaxies, and  in \S\ref{sec_SMBH}
we examine its implications for the SMBH mass - galaxy mass 
relation. We will see that applying the observed 
$M_{\rm BH}$ - $M_{\rm cb}$ relation to the modeled galaxies 
reproduces well the observed relation between $M_{\rm BH}$ and 
the total stellar mass of the host galaxies. This suggests that 
the co-evolution of SMBH is predominantly with classical 
bulges formed through mergers, while secular 
evolution does not play a major role in the growth 
of SMBH. We study the implications of our results 
for AGN feedback in \S\ref{sec_feedback}, and 
summarize in \S\ref{sec_discussion}.    

\section{Classical Bulges}
\label{sec_model}

In this section we provide a very brief description of the model of 
\citet{LuZ14a,LuZ14b}, adopted here to make predictions for the masses 
of classical bulges in galaxies, referring the reader to the 
original papers for details. 

The star formation rate (SFR) of a central galaxy 
in a halo at a given redshift $z$ is assumed to depend 
only on the virial mass of the host halo, $M_{\rm h}(z)$ and $z$. 
Simple broken power laws are adopted for the mass and redshift dependence. 
The build-up of individual dark matter halos is modeled 
using the halo merger tree generator developed by \citet{Parkinson08}. 
This is a Monte-Carlo model based on a modified treatment of the 
extended Press-Schechter formalism and calibrated with 
$N$-body simulations \citep[see][]{Cole08}.
The stellar masses and luminosities  
of the galaxies in a halo and in its progenitors
are obtained by going through its merger tree and integrating 
the SFR with an assumed stellar evolution model and a stellar
IMF.  The exact form of the mass and redshift dependence is 
inferred from a set of observational constraints using
a Bayesian technique.

In \citet{LuZ14a, LuZ14b}, two model families are considered
in detail. The first one, labelled as Model II in these papers, 
is constrained by the observed stellar mass functions (SMF) in 
the redshift range $0$ to $4$.
Model II represents a class of `Slow Evolution' models
studied in the literature
\citep{Behroozi13a, Behroozi13b, Yang12, Yang13}:
virtually all stars form in dark matter halos in a narrow range of halo mass
$10^{11} \Msunh \la M_{\rm h} \la 10^{12} \Msunh$ over 
a large range of redshift. For this model, galaxy-galaxy merger 
is a significant channel of mass assembly only for the most massive 
galaxies in halos with $M_{\rm h} > 10^{13}\Msun$; 
galaxies with lower masses acquire their stellar mass mainly
through {\it in situ} star formation.

The second one, the fiducial model and labelled as Model III, 
is constrained not only by the SMF, but also the luminosity 
function of cluster galaxies \citep{Popesso06}. The latter shows
a steep upturn at the faint end, which provides an important 
constraint on star formation histories in low-mass halos. 
For halos more massive than $10^{11}\Msun$, Model III and Model II
are consistent with each other. In halos with $M_{\rm h} <10^{11}\Msun$, 
however, Model III predicts more efficient star formation at $z>2$.
This efficient star formation phase during the early epoch leads
to some interesting predictions that are supported by recent
observations, such as the presence of a significant old 
stellar population in dwarf galaxies \citep{Weisz11} and 
the steep slopes of the galaxy stellar mass 
and star formation rate functions \citep{Smit12}.  
The efficient star formation at the early epoch 
is accompanied by frequent galaxy-galaxy mergers, 
as shown in \citet[][see the lower left panel of figure 4 in
their paper]{LuZ14b}, which 
is expected to leave other important imprints in the 
galaxy population.

In this paper, both models are used to study the formation of 
classical bulges and the connection between SMBH and their host galaxies.
Model II is also included to show how star formation at high $z$
in low-mass halos can affect our model predictions.

\begin{figure*}
\plotone{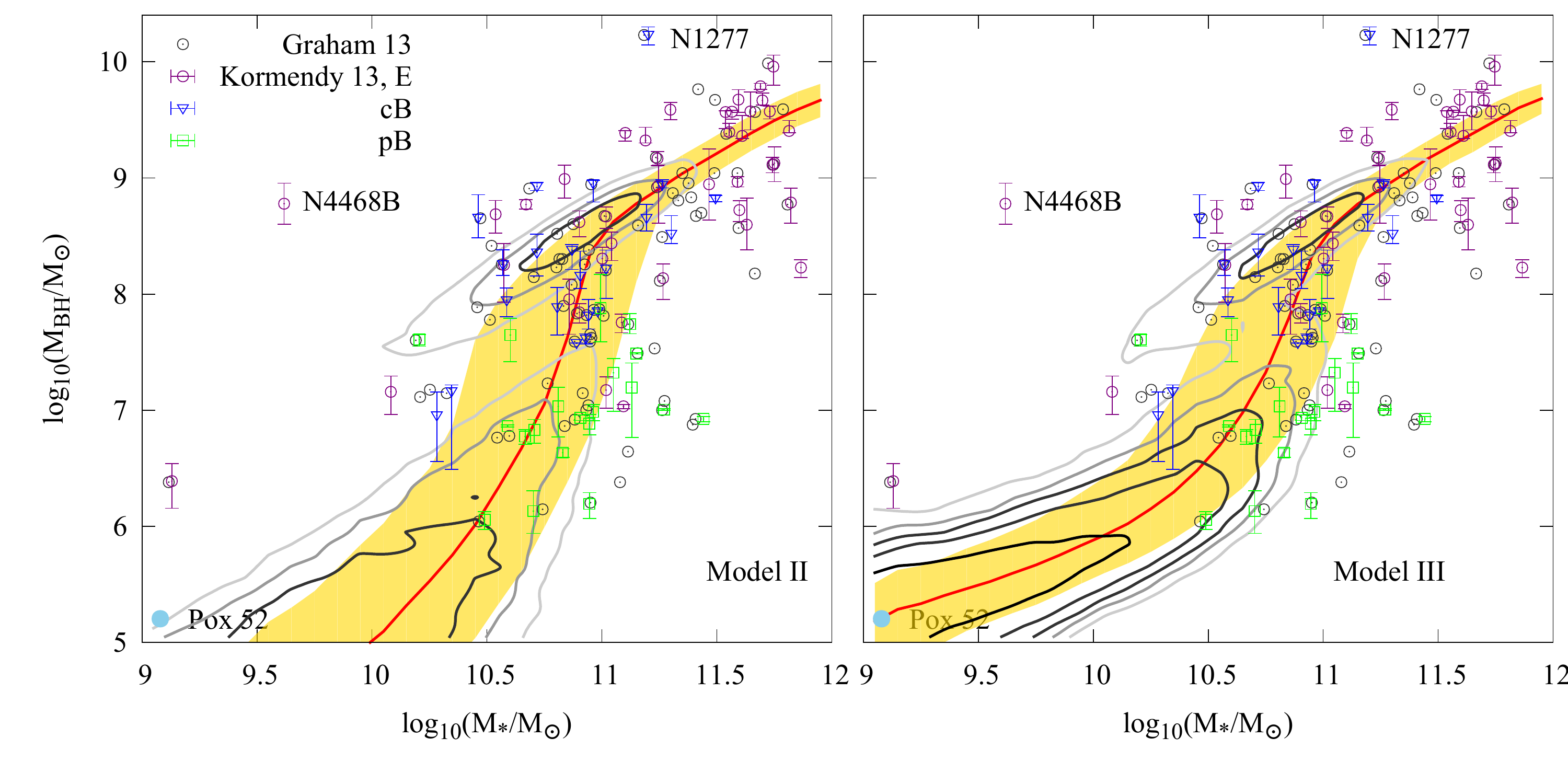}
\caption{
Relation between the SMBH mass and the 
total galaxy stellar mass predicted by Model II (left) 
and Model III (right). The red line is the median of the model 
prediction in each stellar mass bin and the yellow band 
encompasses $90\%$ of the galaxies in the corresponding bins.
For comparison, iso-density contours in the 
$M_{\rm BH}$ - $M_\star$ plane are also shown. 
The colored data points are compiled by \citet{Kormendy13}.  
The two outliers, N4468B and N1277, probably have  
their stellar masses significantly 
reduced due to strong stripping.  
The black open circles are data from
\citet{Graham13}.
The light blue point at the bottom left corner is 
the dwarf Seyfert 1 galaxy Pox 52 \citep{Thornton08}.
}
\label{fig_MbhMs}
\end{figure*}

Following \citet{LuZ14b}, we assume that the stellar disks of 
galaxies form only through {\it in situ} star formation, and that a major merger 
can transform a disk into a classical bulge. 
Thus the mass of the stellar disk, $M_{\rm d}$,  
is simply the total mass of  stars formed {\it in situ} 
after the last major merger, while the mass of the classical 
bulge, $M_{\rm cb}$, is just the difference between the total 
stellar mass, $M_\star$, and $M_{\rm d}$. Figure~\ref{fig_bulge} 
shows the relation between $M_{\rm cb}$ and $M_\star$ 
for central galaxies predicted by Model II (left) and 
Model III (right). 
Owing to frequent major mergers,
the most massive galaxies are completely dominated by the classical 
bulge components. Both models predict a strong transition 
at $M_\star\sim 10^{11}\Msun$. For galaxies in the mass range
$2\times 10^{10}\Msun$ to $10^{11}\Msun$, {\it in-situ}
star formation begins to quench at the present time 
while major mergers are sparse but not negligible. 
Consequently, these galaxies experience either one 
or no major merger in the recent past, so that they are either 
dominated by a classical bulge or remains disk dominated, 
producing the strong bi-modality in the $M_{\rm cb}$ - $M_\star$ relation 
shown in Figure\,\ref{fig_bulge}. 
For dwarf galaxies with $M_\star\ll 10^{10}\Msun$,  
Model II predicts that the mass in the classical bulge 
component is much smaller than the total stellar mass 
of the galaxy. Model III, however, predicts much larger masses 
for the classical bulge components, because of the enhanced star 
formation in low-mass galaxies and the more frequent 
galaxy-galaxy major mergers at $z>2$
\citep[][see the lower left panel of figure 4 in the paper]{LuZ14b}. 

\section{Implications for supermassive black holes}
\label{sec_SMBH}

Given that all galaxies are expected to contain classical bulges 
formed through mergers of galaxies, we test the hypothesis 
that SMBHs are only due to classical bulges. 
The correlation between SMBHs and classical bulges from \citet{Kormendy13} is
\begin{equation}
 \label{eq:Mbh-Mcb}
 \frac{M_{\rm BH}}{10^{9}\Msun} = 
 \left(0.49^{+0.06}_{-0.05}\right) 
 \left(\frac{M_{\rm cb}}{10^{11}\Msun}\right)^{1.16\pm0.08}\,.
\end{equation} 
The intrinsic scatter in $M_{\rm BH}$ is about $0.29$ dex at fixed $M_{\rm cb}$.
We apply this relation to the classic bulges obtained in the last section
to predict a SMBH mass for each model galaxy. 
The red curve in Figure\,\ref{fig_MbhMs} shows the mean of the 
predicted SMBH mass as a function of the total stellar mass, 
and the iso-density contours show the distribution of galaxies in 
the $M_{\star} - M_{\rm BH}$ plane.  
Results are presented for both Model II (left) and Model III (right). 
The predicted relations are compared with the observational 
data compiled by \citet{Kormendy13}, shown as colored  
points with different styles representing different systems,
as well as the catalog from \citet{Graham13}, shown as
black open circles.
The dwarf Seyfert 1 galaxy Pox 52 \citep{Thornton08} is also plotted
(the solid light blue points).

\begin{figure}
\plotone{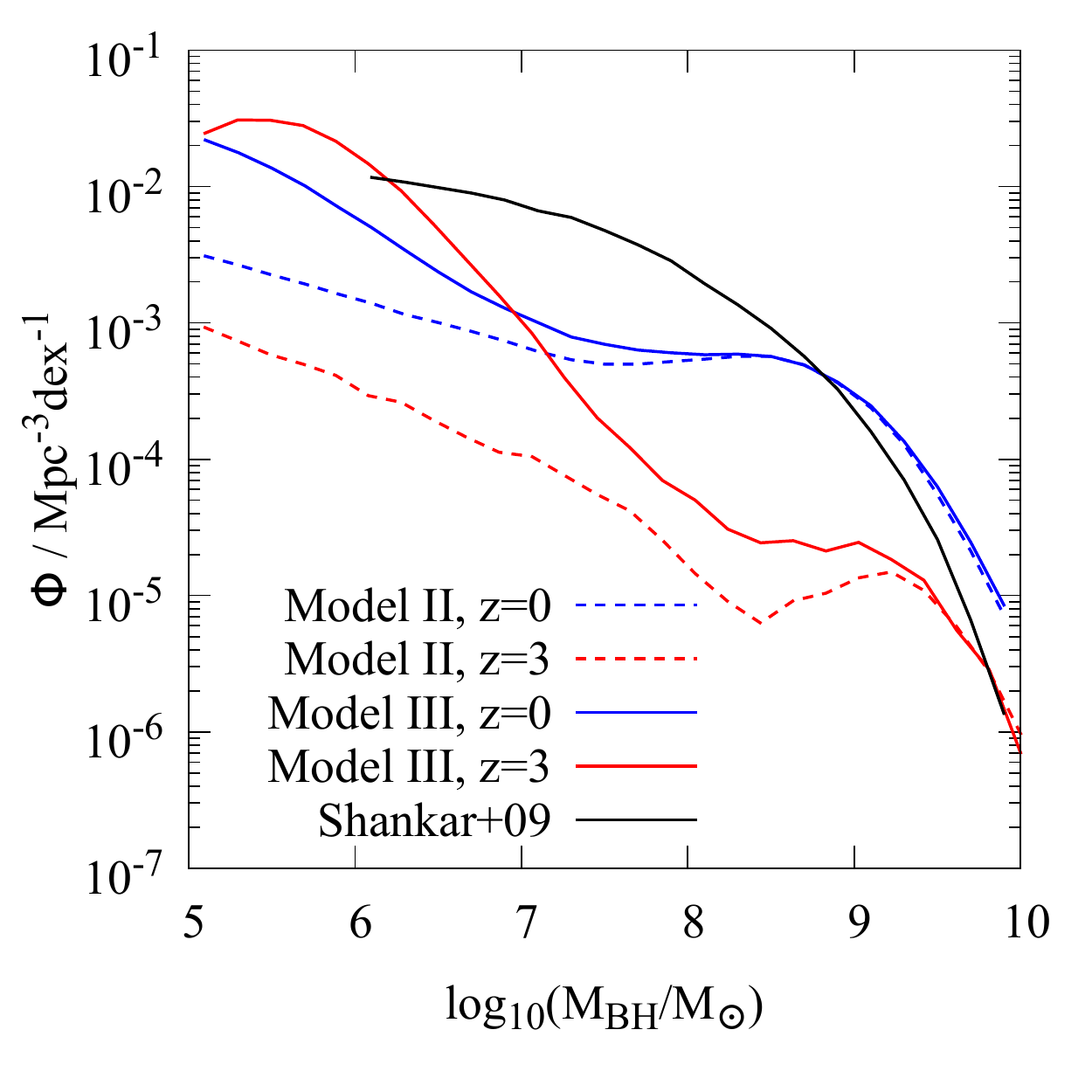}
\caption{SMBH mass function predicted 
by Model II (colored dashed lines) and Model III (colored solid lines) at 
$z=0$ (blue) and $z=3$ (red). 
The black line is the mass function of \citet{Shankar09}.}
\label{fig_bhmf}
\end{figure}

Both models reproduce the strong transition at 
$M_\star=10^{10.5}$ - $10^{11}\Msun$, below which the value 
of $M_{\rm BH}$ drops rapidly with decreasing $M_\star$.
In the observational data, the transition occurs at a place 
where pseudo-bulges start to take over classical bulges. 
In the model predictions, the transition is due to the rapid 
decline of classical bulge mass with decreasing $M_\star$, 
as shown in Fig.\,\ref{fig_bulge}. For galaxies with  
$M_\star<10^{10.5}\Msun$,  Model II predicts a rapid decline 
of SMBH mass with decreasing stellar mass, 
$M_{\rm BH} \propto M_\star ^{2.5}$, with large scatter. 
In contrast, Model III predicts a new $M_{\rm BH} \propto M_\star$
sequence, but with the amplitude dropped by a factor of 
$\sim 50$ relative to that at the massive end. 
At the moment, the small number of observational data points
appear to be better matched by Model III, but no reliable conclusion 
can be reached yet without more data and a better understanding 
of observational selection effects.
It is clear, though, that Model III, which is favored by a large set 
of observational data, does predicts that many galaxies with 
sub-Galactic masses host SMBH with 
$M_{\rm BH}=10^5$ - $10^6\Msun$.   

Figure~\ref{fig_bhmf} shows the SMBH mass functions predicted 
by Model II (dashed curves) and Model III (solid curves) 
at $z \approx 0$ (blue) and $z \approx 3$ (red).
For $z=3$ we have adopted a $M_{\rm BH}/M_{\rm cb}$ ratio
that is three times as large as that for $z=0$ \citep{Kormendy13}.    
For $M_{\rm BH} > 10^{8}\Msun$ and $z=0$, the predictions of the two 
models are almost identical. For $M_{\rm BH} < 10^{7}\Msun$, 
however, 
the mass function predicted by Model III is much steeper than
that predicted by Model II, particularly at high $z$.
According to Model III, most SMBH in this mass range,
which are hosted by galaxies with sub-Galactic masses, 
have already in place at $z\approx3$. They formed
together with their host galaxies in dwarf halos with 
$M_{\rm h}\la 10^{11}\Msun$ at $z>2$, where star formation 
is efficient and gas-rich mergers of galaxies are still frequent 
\citep{LuZ14b}. 
The consequence of such SMBH formation for the evolution of 
their host galaxies is discussed in the next section. 

For comparison, we also plot the SMBH mass function estimated
by \citet{Shankar09}, which is widely adopted in the literature.
Without distinguishing between pseudo-bulges and classic bulges,
their estimates assumed that all late-type galaxies 
have a bulge-to-total light ratio of $\approx 0.27$ \citep{Fukugita98}.
This ratio is about $7$ times larger than our prediction for
classical bulges, and is also the reason why their mass function 
is much higher than ours at $M_{\rm BH}\sim 10^{7.5}\Msun$.

\section{AGN feedbck from low-mass galaxies}
\label{sec_feedback}

As a SMBH grows, the power of the energy output can be 
written as 
\begin{equation}
{d\, E \over d\, t} = \epsilon {\dot M}_{\rm BH} c^2\,,
\end{equation}
where $\epsilon$ is an efficiency factor. The total energy 
output is
\begin{equation}
E={\overline \epsilon} M_{\rm BH} c^2\,,
\end{equation}
where ${\overline \epsilon}$ is the mean efficiency.  
Since in our model $M_{\rm BH}$  is roughly 
proportional to $M_{\rm cb}$, and the 
energy feedback from star formation is also $\propto M_{\rm cb}$, 
the above equation (and our modeling below) also applies even 
if the feedback effect of classical bulge stars is taken into account.

Suppose a fraction $f_\epsilon$ of
this energy output is eventually transferred to 
and retained in a total amount of gas of mass 
$M_{\rm gas}$ and write this mass in terms 
of the mass of the host halo: 
$M_{\rm gas}=\lambda f_{\rm B} M_{\rm h}$, with $\lambda$
a constant loading factor and $f_{\rm B}\approx 0.17$ the universal
baryon fraction. The effective temperature of the 
gas due to the energy transfer can then be written as 
\begin{equation}
{k T \lambda f_{\rm B} M_{\rm h}\over \mu}
=f_\epsilon {\overline \epsilon} M_{\rm BH} c^2\,,
\end{equation}
where $\mu$ is the mean molecular weight of the gas.
This can be written in a more useful form:
\begin{equation}
{k T \over \mu}
={f_\epsilon\overline \epsilon c^2\over \lambda f_{\rm B}} 
\left({M_{\rm BH}\over M_\star}\right)  
\left({M_\star\over M_{\rm h}}\right)\,.
\end{equation}
As shown in \citet{LuZ14b}, for dwarf galaxies with stellar masses
between $10^{8}$ and $10^{9}\Msun$ at $z\sim 3$, 
which were the progenitors of present day
sub-Galactic galaxies ($10^9-10^{10}\Msun$),
$M_\star/M_{\rm h}\approx 10^{-2.5}$ 
(see the bottom right panel in figure 3 of their paper). 
These progenitors are typically bulge dominated because of 
galaxy-galaxy major mergers. If we adopt a bulge to total ratio
of $1/2$ and scale Eq.\,(\ref{eq:Mbh-Mcb}) up by a factor of $3$
for these high-redshift progenitors \citep{Kormendy13},    
we get $M_{\rm BH}/M_\star \approx 0.3\%$.
Assuming $\mu$ to be $0.6$ times the proton mass, and 
${\overline\epsilon}=0.1$ as is usually assumed 
for an AGN in the quasar mode, we have 
\begin{equation}\label{eq_Qheat}
T_6\approx 1.0\times \left({f_\epsilon \over 0.03}\right)
\left({\lambda \over 10}\right)^{-1}\,,
\end{equation}
where $T_6\equiv T/10^6{\rm K}$. 

For the current cosmology, the age of the universe 
at $z>1$ is roughly 
\begin{equation}
t \approx 3.0\times \left({1+z\over 3}\right)^{-3/2}
{\rm Gyr}\,,
\end{equation}
and the cooling time of the gas medium with an over-density 
$\delta$ and temperature $T$ at $z$ is
\begin{equation}
t_{\rm cool} \approx 3.0\times \left({1+z\over 3}\right)^{-3}
\Lambda_{-23}^{-1} T_6 \left({1+\delta\over 60}\right)^{-1}
{\rm Gyr}\,,
\end{equation}
where $\Lambda_{-23}$ is the cooling function in units 
of $10^{-23}{\rm erg\, s^{-1}{\rm cm}^3}$.
Setting $t_{\rm cool}=t$ and using the fact that
$\Lambda_{-23} \approx T_6^{-1}$ for a low-metallicity 
gas with temperature in the range
$10^{5}$ - $10^6{\rm K}$, we get
\begin{equation}\label{eq_cool}
T_6 \approx 1.0\times 
\left({1+z\over 3}\right)^{3/4}
\left({1+\delta\over 60}\right)^{1/2}\,.
\end{equation}
This temperature corresponds to a specific entropy
\begin{eqnarray}\label{entropy}
S&\equiv& {kT \over n_{\rm e}^{2/3}}
\approx 15.0 \times T_6 \left({1+\delta\over 60}\right)^{-2/3}
\left({1+z\over 3}\right)^{-2} {\rm keV\,cm^2}\nonumber\\
 &\approx& 15.0\times \left({1+\delta\over 60}\right)^{-1/6}
\left({1+z\over 3}\right)^{-5/4} {\rm keV\,cm^2}\,.
\end{eqnarray}
Note that $S$ depends only weakly on $\delta$ at a given $z$. 

If the medium is heated above the temperature
given by Eq.\,(\ref{eq_cool}), gas cooling will be suppressed.
Thus, for the AGN feedback to have a significant impact 
on subsequent galaxy formation and evolution, 
the temperature given by Eq.\,(\ref{eq_Qheat})
should be higher than that given by Eq.\,(\ref{eq_cool}).
This gives a constraint on $\lambda$, 
\begin{equation}\label{eq_critz}
{\lambda \over 10}
\lesssim 
\left({f_\epsilon \over 0.03}\right)
\left({1+z \over 3}\right)^{-3/4}
\left({1+\delta\over 60}\right)^{1/2}\,.
\end{equation} 
Numerical simulations show that to reproduce the 
observed $M_{\rm BH}$ - galaxy velocity dispersion  
relation requires $5\%$ of the total energy output
be coupled with the surrounding gas
\citep[e.g.][]{diMatteo08}. The energy will propagate
through the host galaxies and the host halos as  blast 
waves or super-bubbles in which radiative cooling is 
expected to be slow because of the low-density of the 
halo gas.  So $f_\epsilon$ is not expected to be 
much lower than a few percent.
The over-density $\delta \approx 60$ is about the 
value appropriate for the exteriors of dark 
matter halos.\footnote{For a singular isothermal halo defined 
so that the mean density within its virial radius is 
$200$ times the mean density of the universe, 
the over-density at the virial radius is $\delta\approx 66$.}    
Finally, as shown in the previous section, most of
the SMBH in the dwarf galaxies formed at $z > 2$.
Setting $f_\epsilon \approx 0.03$, $z \approx 3$ and 
$\delta \approx 60$ gives $\lambda \lesssim 8$. 
As shown in \citet{Zhao09}, a low-mass halo has typically 
increased its mass by a factor of 5  since $z\sim 3$.
Thus, the AGN feedback is, in principle, capable 
of affecting all the gas to be accreted by a halo in its 
subsequent growth.

How might such a feedback proceed?  In general, the impact 
of the feedback from an AGN is expected to first 
affect the gas close to the galaxy. Since the IGM is 
expected to be clumpy and filamentary, the dense part of the medium 
may still be able to cool and circumvent total disruption by 
the wind. This part of the gas can then move against the outflow 
toward the halo center to feed the galaxy and SMBH. 
The growth of a SMBH and star formation may thus go through a 
number of bursts, as the dense clouds are accreted 
episodically. As the local medium is heated
and expands, an negative gradient of entropy 
may develop, producing gas convection and mixing gas 
of different specific entropies.  Eventually a large 
envelope of roughly constant specific entropy may 
develop around each low-mass halo. This process 
continues until most of the gas in the IGM around 
the galaxy has an specific entropy such that its radiative 
cooling time is comparable to the age of the universe. 
As demonstrated above, the likely epoch for this  
to accomplish is at $z \sim 2$ to $3$, and the specific 
entropy is of the order of $10$ - $15\,{\rm keV\,cm^2}$.
The subsequent formation and 
growth of dark matter halos are then in a preheated 
medium where gas accretion and cooling in low-mass halos
can be reduced significantly. This scenario of galaxy 
formation in a preheated medium was first proposed 
in \citet{Mo02}. As shown in 
\citet{Mo05}, \citet{LuY07} and \citet{LuY14}, 
the level of specific entropy predicted here is 
roughly what is needed to suppress 
gas accretion and star formation in low-mass halos 
to match the observed stellar mass and HI-mass 
functions at the low-mass end.    

\section{Discussion}
\label{sec_discussion}

In this paper we have shown that the observed SMBH masses are 
consistent with the assumption that they are directly proportional 
to the mass of the classical bulges of their hosts, even in late-type 
galaxies. In particular, this assumption combined with the 
classical bulge masses reproduces well the rapid decrease of 
$M_{\rm BH}$ with decreasing $M_\star$ at $M_\star\la 10^{11}\Msun$. 
Our Model III, constrained by various observations of galaxies, 
predicts a new low-mass sequence in the  $M_{\rm BH}$ - $M_\star$
relation, where the SMBH mass is roughly proportional to galaxy 
mass, but with the amplitude $\sim 50$ times 
lower than that of the high-mass sequence. 

The assumptions in and consequences of our proposed scenario 
can be tested with more detailed modeling.  
At the moment, it is still unclear whether SMBH in low-mass 
and late-type galaxies form in a similar way as in 
their more massive counterparts. For example, it is unclear 
whether they grow from seeds similar to that for massive 
SMBH, and whether their growth is also dominated 
by quasar modes, although they are predicted to be 
produced by mergers of gas-rich galaxies. 
To answer these questions, observations 
of low-luminosity AGNs at high $z$ and accurate 
determinations of the SMBH masses in low-mass galaxies 
are essential. If our scenario is correct, we expect to 
observe a large number of low-luminosity AGNs at 
$z>2$ and a large number of SMBH with masses in the 
range from $10^5$ to $10^6\Msun$ at $z\sim 0$.
The classical bulges within which these SMBH have formed 
are expected to be compact, not only because their 
host halos at $z>2$ are small, but also because 
major galaxy-galaxy mergers can reduce the angular
momentum of the cold disk gas, making the merger 
remnants even smaller. How such formation is related 
to the observed ultra compact dwarfs \citep[UCD, e.g.][and references therein]{Norris14}
is clearly an interesting and open question.   

In our scenario an implicit assumption is that secular 
evolution of galaxy disks does not play a major role in 
the growth of SMBH. This assumption can be tested by 
observing barred disks, in which secular evolution is on-going, 
to see whether their AGN activities are elevated or not. 
So far observational evidence is negative for such elevation 
\citep[e.g.][]{Kormendy11},  but current data are still sparse.

We have also shown that the formation of SMBH in 
low-mass galaxies provides a new scenario of preventative 
feedback, in which gas is prevented from being accreted 
into a dark matter halo due to preheating. Such a scenario 
is plausible based on our simple arguments, but the details 
need to be worked out. 
For example, how does the AGN-driven outflow propagates into 
the IGM and affect its properties? What is the structure of 
the circum-galactic medium (CGM) produced in this 
way? Will such formation produce a multiphase medium 
\citep{MoMiralda96} and how will it be observed 
in QSO absorption line studies? Can the interaction between 
the gas and dark matter reduce the phase space density 
of dark matter halos, as advocated in \citet{Mo04}? 
We will come back to some of these questions in the 
future.
\section*{Acknowledgements}

We thank Luis Ho and Alister Graham for helpful discussion.  
HJM would like to acknowledge the support of NSF AST-1109354.
\\

\end{document}